\documentclass[aps,prl,showpacs, amsmath,amssymb,twocolumn]{revtex4}

\usepackage{graphicx}
\usepackage{dcolumn}
\usepackage{bm}

\newcommand{\avg}[1]{\langle #1 \rangle}

\begin{document}

\title{Spatial correlations in polydisperse, frictionless two-dimensional packings}
\author{C.B. O'Donovan, M.E. M\"{o}bius}
\affiliation{School of Physics, Trinity College Dublin, Dublin 2, Ireland}

\date{\today}

\begin{abstract}
We investigate next nearest neighbor correlations of the contact number in simulations of polydisperse, frictionless packings in two dimensions. We find that discs with few contacting neighbors are predominantly in contact with discs that have many neighbors and vice versa at all packing fractions. This counter-intuitive result can be explained by drawing a direct analogy to the Aboav-Weaire law in cellular structures. We find an empirical one parameter relation similar to the Aboav-Weaire law that satisfies an exact sum rule constraint. Surprisingly, there are no correlations in the radii between neighboring particles, despite correlations between contact number and radius.
\end{abstract}
\maketitle

Disordered packings of particles are the quintessential model for amorphous materials such as granular packings \cite{vanHecke}, emulsions \cite{Clusel}, wet foams \cite{Bolton} and glass formers \cite{Wyart}. While the contact number distribution and its average near the random close packing density \cite{Durian, Clusel, Bernal, Bolton, Katgert, Ohern, vanHecke} have been extensively studied in these systems, little is known about spatial correlations in the contact network. Various models  that have recently been put forward to predict the density \cite{Makse}, distribution of contact numbers \cite{Clusel} and forces \cite{Tighe} in random close packings implicitly assume the absence of such correlations.

We address this question through simulations of a two-dimensional model system with polydisperse, frictionless soft discs. At the random close packing density $\phi_{c}$, the discs just touch and have an average contact number $\avg{z}$ close to $4$ as required for mechanical stability \cite{Maxwell,Alexander,vanHecke}. Due to disorder the individual contact numbers $z$ are distributed according to some distribution $P(z)$ that is typically non-Gaussian and depends on the polydispersity of the disc size distribution \cite{Clusel}. Here, we investigate whether spatial correlations in the contact network exist. 

In our simulations we find that discs with many contacts favor neighbors with fewer contacts and vice versa. These correlations persist for all packing densities. This result is a direct analogue to the well known Aboav-Weaire law in the field  of cellular structures which states that cells with fewer neighbors are surrounded by cells with many neighbors \cite{Schliecker,bible,Aboav,Weaire,Stavans}. We will show that our results are in excellent agreement with a modified Aboav-Weaire law. Since geometrical constraints in the packing dictate that smaller particles have fewer contacts on average \cite{Clusel}, one may expect similar correlations for the size distribution in the packing, namely that larger particles are surrounded by smaller ones. Surprisingly, we find that the size of the central particle is uncorrelated to the average size of the contacting neighboring particles.

We simulate the disordered packings by using Durian's soft disc model \cite{Durian}, as implemented by Langlois et al. \cite{Langlois}. The discs have a harmonic repulsion proportional to their overlap and experience viscous tangential drag. We use $1500$ polydisperse discs whose size is normally distributed around the mean $\avg{r}$ with a variance $\sigma_2=(0.304 \avg{r})^2$. The polydispersity allows us to access a wide range of contact numbers which is important to measure next nearest neighbor correlations. The discs are randomly placed in a periodic box at low packing fraction and then allowed to relax while their radii are slowly increased. The simulation terminates when the total elastic energy due to overlaps reaches a steady state at a predefined packing fraction. For each packing density, up to $10$ different packings are created to increase the statistics of our correlation measures. Upon reaching equilibrium, discs with fewer than three contacts (rattlers) are removed for the analysis of the contact network but are accounted for in the packing fraction. Contacts are defined as overlaps between discs.

\begin{figure}
\begin{center}
\includegraphics[width=3.2in]{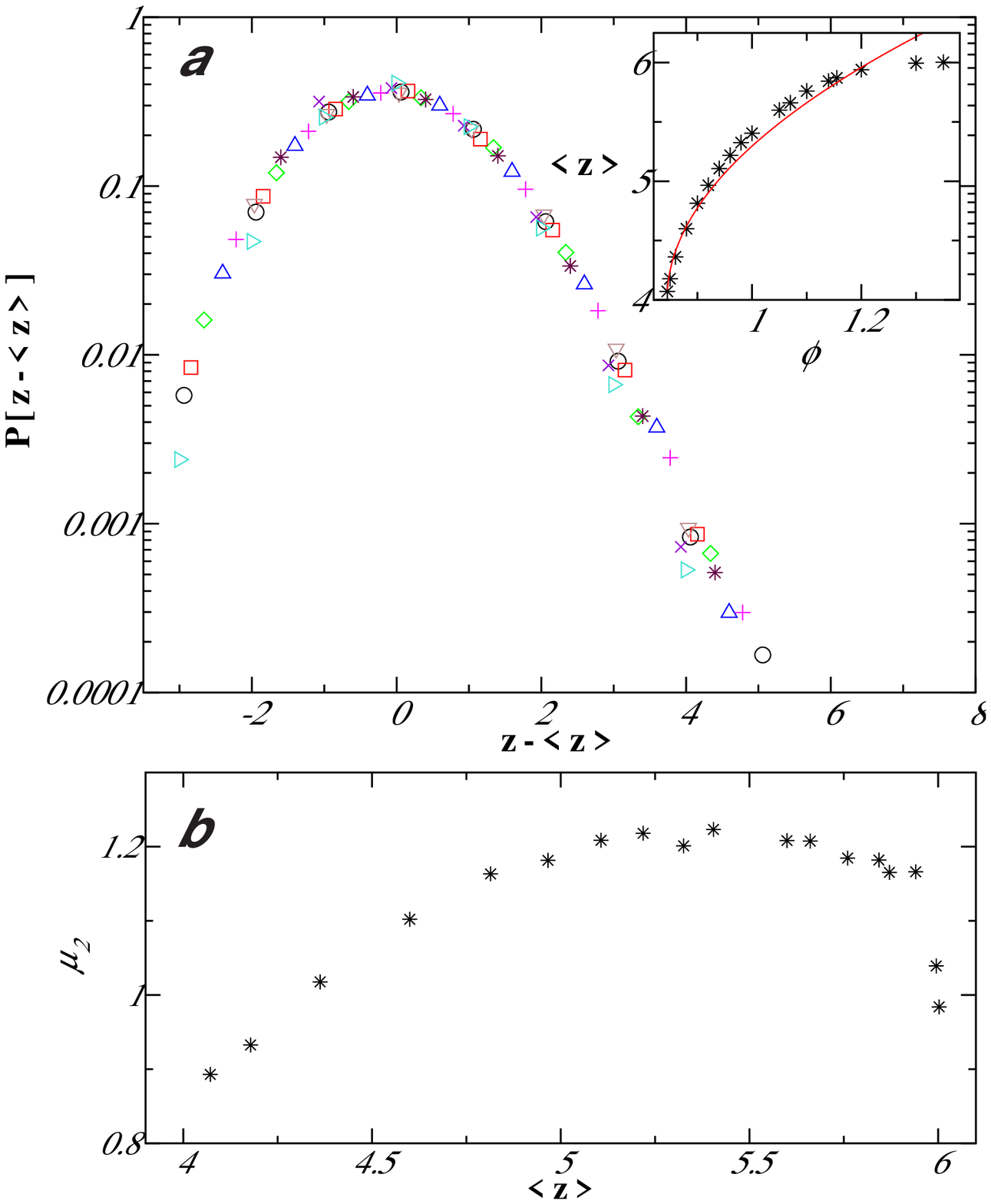}
\caption{\label{fig1} (a) The contact number distribution $P(z-\avg{z})$ for different packing fractions $\phi$: $(\times)$ $0.845$; $(\ast)$ $0.88$; $(\triangledown)$ $0.92$;  $(+)$ $0.96$; $(\vartriangle)$ $1.0$; $(\lozenge)$ $1.07$; $(\square)$ $1.14$; $(\bigcirc)$ $1.2$; $(\vartriangleright)$ $1.35$. The inset shows the variation of the average contact number $\avg{z}$ with packing fraction $\phi$. The red line corresponds to the square root scaling fit: $\avg{z}=4+z_0\sqrt{\phi-\phi_{c}}$, where $z_0=3.29$. (b) The variance of $P(z)$, $\mu_2$, versus $\avg{z}$.}
\end{center}
\end{figure}

We study packings which range from the random close packing density $\phi_{c}$ up to $\phi=1.35$. Note that the overlap area between bubbles is counted twice in the calculation of $\phi$ in line with previous simulations of packings \cite{vanHecke}. In our simulations we find $\phi_c=0.845$ and the corresponding average contact number $\avg{z}=4.07$, which is close to the isostatic prediction $\avg{z}=4$ \cite{vanHecke}. As shown in the inset of Fig.\ref{fig1}, $\avg{z}$ increases approximately as $4+3.29\sqrt{\phi-\phi_{c}}$ close to the isostatic point, which is consistent with previous results \cite{Ohern, Katgert}.
We study packings up to $\phi=1.35$, where the average contact number reaches $\avg{z}=6$.

Figure \ref{fig1}(a)-\ref{fig1}(b) shows  the distribution of the relative contact number, $P(z-\avg{z})$, for different packing fractions and their respective variance $\mu_2=\avg{z^2}-\avg{z}^2$, where $\avg{z^i}=\sum_z z^iP(z)$. The shape of the distribution, which is not Gaussian, is independent of the packing fraction as evidenced by the collapse of $P(z-\avg{z})$ onto a master curve, while $\mu_2$ varies only slightly.

Next, we address the main result of our work -- the correlations in the contact network of the packings at different densities. Given a disordered packing of frictionless discs with a certain global average contact number $\avg{z}$, are the \emph{local} contact numbers of neighboring particles correlated? Here, we define neighbors to be discs in contact, i.e. discs that overlap. In order to quantify nearest neighbor correlations, we measure $m(z)$, which is the average contact number of the neighbors of a disc with contact number $z$.

This analysis is analogous to the pioneering work of Aboav \cite{Aboav} on polycrystals, where the same analysis was performed for the coordination number of cells in a cellular structure. He found that many-sided cells are surrounded by few-sided cells on average and vice versa. It has been noted by Weaire \cite{Weaire}, that $m$ obeys an exact sum rule which is independent of dimensionality.:
\begin{equation}\label{sumrule1}
\sum_{z}mzP(z)=\sum_{z} z^2 P(z),
\end{equation}
 where $z$ is the coordination number of the cells. This sum rule is based on a counting argument and one can show that the sum on the left hand side of the equation counts cells with $z$ neighbors $z$ times. In the absence of correlations in the coordination number between neighboring cells, $m(z)$ is simply a constant ($\equiv\overline{m}$). It follows from the sum rule that in this case $\overline{m}=\avg{z}+\mu_2/\avg{z}$ \cite{bible}. In the context of cellular structures, this is referred to as a topological gas, although its existence is disputed \cite{Schliecker}. The Aboav-Weaire relation is a solution of the sum rule:
 \begin{equation}\label{AWlaw}
   m=\avg{z}-a+(\avg{z}a+\mu_2)/z,
 \end{equation}
where $a$ is an empirical parameter . For a topological gas, $a=-\mu_2/\avg{z}$, but in a natural cellular structures such as dry foam $a\approx 1$ \cite{bible,Stavans}. Therefore, many sided cells have few-sided neighbors and vice versa. One can interpret this anti-correlation as a partial screening of the topological charge $(z-\avg{z})$ by its nearest neighbors whose combined charge is $z(m-\avg{z})$. Most 2D cellular structures, such as polycrystals and dry foams, obey this relation well \cite{Aboav,Stavans,Schliecker}, with the notable exception of random Voronoi tessellations that exhibit deviations from this rule \cite{Hilhorst}.

The counting argument that leads to the sum rule (Eq. (\ref{sumrule1})) was originally developed for cellular structures but holds equally well for neighbors in a contact network of a disordered packing. The main difference is that in 2D cellular structures with threefold vertices, $\avg{z}=6$ \cite{bible}, while frictionless packings in two dimensions have an average contact number $\avg{z}$ greater than or equal to $4$ depending on the packing fraction \cite{Ohern,vanHecke}.

\begin{figure}
\begin{center}
\includegraphics[width=3.2in]{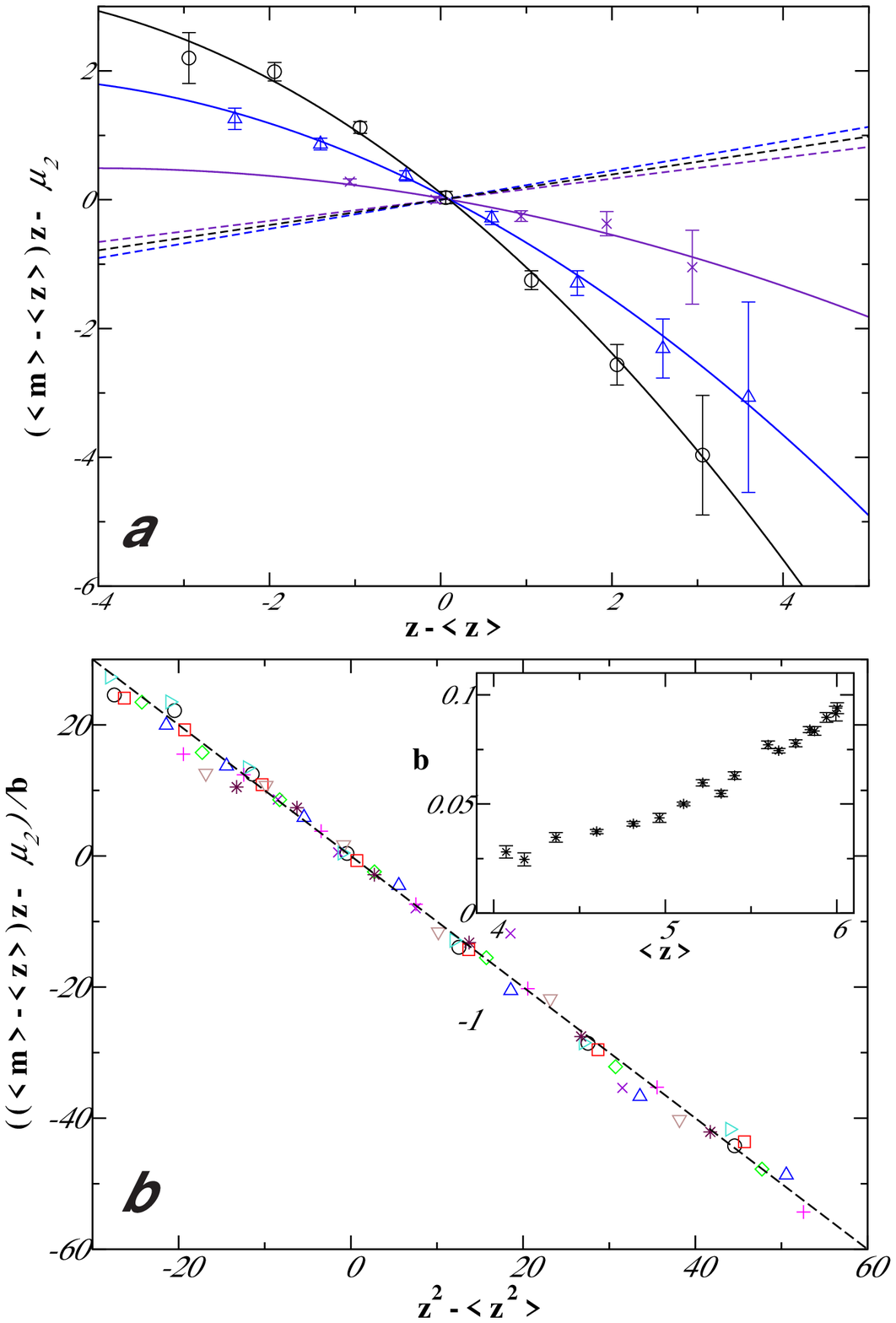}
\caption{\label{fig2} Nearest neighbor correlations of the contact number. (a) $(m-\avg{z})z-\mu_2$  versus $(z-\avg{z})$ for three different densities: $(\times)$ $0.854$; $(\bigtriangleup)$ $1.0$; $(\bigcirc)$ $1.2$. The error bars are standard deviations from the mean. The lines are fits to Eq. (\ref{AW2}). The dotted lines correspond to the uncorrelated prediction $a=-\mu_2/\avg{z}$. (b) $[(m-\avg{z})z-\mu_2]/b$ versus $(z^2-\avg{z^2})$ for all densities (same symbols as in Fig.\ref{fig1}(a)). The dotted line corresponds to the slope $-1$. The inset shows the fit parameter $b$ as a function of $\avg{z}$.}
\end{center}
\end{figure}

The results for $m$ are shown in Fig.\ref{fig2}(a), where we plot $(m-\avg{z})z-\mu_2$ versus $(z-\avg{z})$ for three different packing fractions. For the Aboav-Weaire law (Eq. (\ref{AWlaw})) to hold we expect the data to follow a line with slope $-a$. Clearly, the data does not follow the uncorrelated prediction $a=-\mu_2/\avg{z}$, instead we observe spatial correlations: discs with few contacts are surrounded by discs with many contacts and vice versa.
Another key result are the deviations from purely linear behavior, especially at higher packing fractions.

In order to account for this non-linearity, we expand the Aboav-Weaire law in terms of the moments of the contact number distribution such that it still satisfies the sum rule
\begin{equation}\label{expand}
(m-\avg{z})z-\mu_2=-\sum_{i=1}^{\infty} c_i(z^{i}-\avg{z^{i}}),
\end{equation}
where the $c_i$'s are arbitrary constants.
If $c_i=0$ for $i>1$, one recovers the usual Aboav-Weaire law with $c_1=a$. In order to fit our data it proved sufficient to only make $c_2$ non-zero, which leads to
\begin{equation}\label{AW2}
m=\avg{z}-bz+\frac{\mu_2(1+b)+b\avg{z}^2}{z},
\end{equation}
where $b=c_2$. This is a one parameter fit, similar to the Aboav-Weaire law. However, $mz$ is now quadratic in $z$, instead of linear. As shown in Fig.\ref{fig2}(b), Eq. (\ref{AW2}) captures the non-linearity well and leads to a much improved fit compared to Eq.(\ref{AWlaw}). Including higher order terms in the expansion (Eq. (\ref{expand})) does not improve the fit significantly. The inset of Fig.\ref{fig2}(b) shows the decrease of the parameter $b$ with $\avg{z}$, which means that the screening of topological charge decreases as the isostatic point is approached.

We would like to stress that the analogy between correlations in the coordination number in disordered frictionless packings and cellular structures is not obvious, since these systems are governed by different local and global constraints. Although polydisperse packings can be tesselated into a cellular structure \cite{Clusel}, not all faces of a cell correspond to contacts, therefore the existence of correlations in packings does not follow naturally from similar correlations in disordered cellular structures.

\begin{figure}
\begin{center}
\includegraphics[width=3.2in]{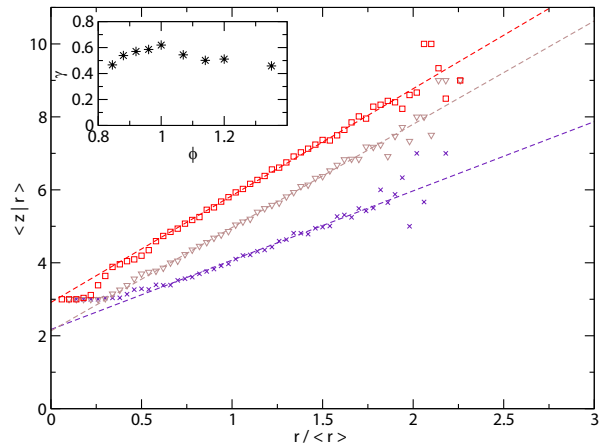}
\caption{\label{fig3} The average contact number $\avg{z|r}$ for a given particle radius $r$ at different packing fractions $\phi$: $(\times)$ $0.845$; $(\triangledown)$ $0.92$; $(\square)$ $1.14$. Lines are fits to Eq.(\ref{lewis}). The inset shows the fit parameter $\gamma$ as a function of $\phi$.}
\end{center}
\end{figure}

There is also a correlation between contact number and particle size in polydisperse packings \cite{Clusel}. Larger particles have more contacts on average, since one may fit more particles around them on average. Figure \ref{fig3} shows the average contact number of a particle with radius $r$, $\avg{z|r}=\sum_z z P(z|r)$, where $P(z|r)$ is the conditional probability of a particle of radius $r$ to have a contact number $z$. It is well described by a linear relation
\begin{equation}\label{lewis}
\avg{z|r}=\avg{z}(1+\gamma(r-\avg{r})).
\end{equation}
Since $\avg{z|r}$ is constrained by the equality $\int_0^{\infty}\avg{z|r}P(r)dr=\avg{z}$, there is only one empirical fit parameter $\gamma$, which varies little with $\phi$ (Fig.\ref{fig3} inset). However, this linear relationship does break down at low $r$, since $\avg{z|r}\ge 3$. A similar result exists in the field of cellular structures and is known as Lewis' law \cite{Lewis,Schliecker}.

Given the nearest neighbor correlations in the coordination number (Eq. (\ref{AW2})) and the correlation between size and contact number (Eq. (\ref{lewis})), are smaller particles surrounded by larger ones, similar to observations in cellular structures \cite{Sire} ?

\begin{figure}[t]
\begin{center}
\includegraphics[width=3.2in]{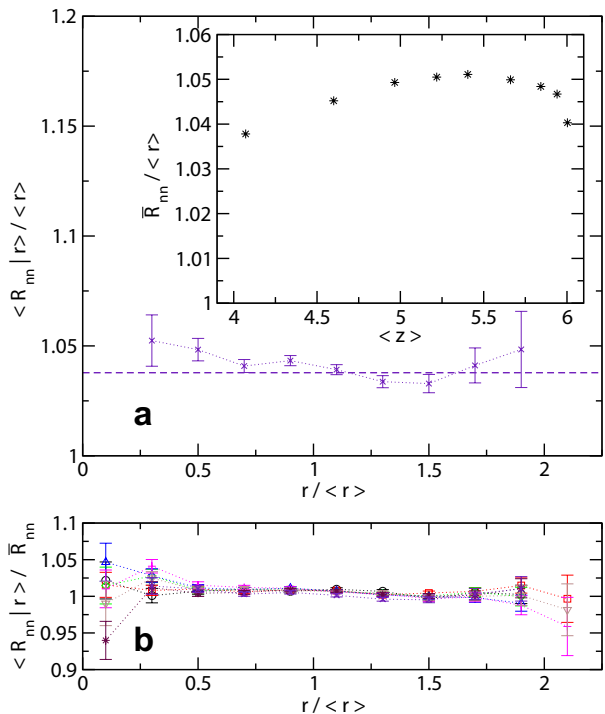}
\caption{\label{fig4} Nearest neighbor correlations of the radii (a) $\avg{R_{nn}|r}$ versus $r/\avg{r}$ for $\phi=0.845$. The dotted line corresponds to the uncorrelated prediction $\overline{R_{nn}}$ (Eq.(\ref{Rnn})). The inset shows $\overline{R_{nn}}$ versus $\avg{z}$. (b) $\overline{R_{nn}} / \avg{R_{nn}|r}$ versus $r/\avg{r}$ for all densities (same symbols as in Fig.\ref{fig1}(a)).}
\end{center}
\end{figure}

In order to study radii correlations, we measured $\avg{R_{nn}|r}$, which is the average radius of neighboring discs in contact with a disc of radius $r$. Figure \ref{fig4} shows $\avg{R_{nn}|r}/\avg{r}$ versus $r/\avg{r}$ for $\phi_c$. No correlations are apparent and the result agrees well with the uncorrelated prediction $\overline{R_{nn}}$ which is discussed below.

The counting argument that leads to the Weaire sum rule for the coordination number can also be applied for the radii.
Analogously, the average radius of the neighbors $\avg{R_{nn}|r}$ needs to satisfy the following relation 
\begin{equation}
  \int_{0}^{\infty} \avg{R_{nn}|r} \avg{z|r} P(r) dr = \int_{0}^{\infty} r \avg{z|r} P(r) dr\\ \label{rnnr}.
\end{equation}
The left hand side of the equation amounts to an integral over the disc radii $r$ weighted by $\avg{z|r}$. In the absence of correlations $\avg{R_{nn}|r}$ is a constant $(\equiv \overline{R_{nn}})$, and we have
\begin{equation}\label{Rnn}
  \overline{R_{nn}}=\frac{\int_{0}^{\infty} r \avg{z|r} P(r) dr}{\int_{0}^{\infty} \avg{z|r} P(r) dr}=\frac{\int_{0}^{\infty} r \avg{z|r} P(r) dr}{\avg{z}}.
\end{equation}
Substituting the empirical relation for $\avg{z|r}$ (Eq.(\ref{lewis})), we find that $\overline{R_{nn}}=\avg{r}(1+\gamma \sigma_2/\avg{r}^2)$, which is slightly larger than $\avg{r}$ and varies little with $\phi$ (inset Fig.\ref{fig4}(a)). At $\phi_c$, we obtain $\overline{R_{nn}}=1.042\avg{r}$, which is consistent with our results from Fig.\ref{fig4}(a). At higher packing fractions shown in Fig.\ref{fig4}(b), $\avg{R_{nn}|r}/\overline{R_{nn}}$ remains constant and close to $1$. Only for high and low $r$, slight deviations due to low statistics are observed.

While the absence of correlations in the radii for neighboring particles may be expected given our preparation procedure where particles are placed in the box at random, it is surprising in the light of the two correlations we have measured. Namely, the correlations between contact numbers of neighboring particles (Eq. (\ref{AW2})) and the correlation between size and contact number (Eq. (\ref{lewis})). The reason for this counter-intuitive result is that the relationship between the average contact number $m$ and the corresponding average radius $R_{nn}$ does not follow the linear relation (Eq. (\ref{lewis})) for a single disc \cite{ODonovan}.


Although we have only shown results for packings with normally distributed radii, similar correlations are observed for other polydispersities such as bidisperse distributions \cite{ODonovan}.

In conclusion, we studied polydisperse, frictionless packings at various packing densities. We find that discs with many contacts are surrounded by discs with few contacts and vice versa. As the isostatic point is approached, the screening of topological charges becomes weaker but does not vanish and is well described by a modified Aboav-Weaire law. This result is a direct analogue the topological screening observed in cellular structures. Nevertheless, the physical origin of the screening parameter $b$ remains unclear much like the $a$ parameter in the Aboav-Weaire law \cite{bible}.

We want to emphasize that the counting argument that leads to the sum rule for the contact number and radii are valid in any dimension. Therefore, one may expect similar correlations in three dimensional packings as well as in frictional packings \cite{Henkes}. It remains to be seen whether these correlations depend on the preparation history of the packing, which is known to have an influence on $\phi_c$ \cite{Chaudhuri}. 

{\em Acknowledgements ---}
C.B. O'D. acknowledges funding from the School of Physics, TCD. The authors are grateful for stimulating discussions with D. Weaire, S. Hutzler and S. Henkes. This work made use of computational facilities provided by the Trinity Centre for High Performance Computing.


\end{document}